\documentclass[submission,Phys]{SciPost}

\usepackage{amssymb}
\usepackage{amsmath}
\usepackage{xspace}
\usepackage{color}
\usepackage{dsfont}

\newcommand{\bvec}[1]{\mbox{\boldmath $#1$}}
\newcommand{\trans}{$\leftrightharpoons$}

\newcommand{\be}{\begin{equation}}
\newcommand{\ee}{\end{equation}}
\newcommand{\bea}{\begin{eqnarray}}
\newcommand{\eea}{\end{eqnarray}}

\def\nn{\nonumber\\}
\def\fr#1{(\ref{#1})}

\begin{document} 

\begin{center}
 {\Large \textbf{Exotic criticality in the dimerized spin-1 $XXZ$ chain}}
 \\
 {\Large \textbf{with single-ion anisotropy}}
\end{center}

\begin{center}
 Satoshi~Ejima\textsuperscript{1*},
 Tomoki~Yamaguchi\textsuperscript{2},
 Fabian~H.~L.~Essler\textsuperscript{3},
 Florian~Lange\textsuperscript{1},
 Yukinori~Ohta\textsuperscript{2},
 Holger~Fehske\textsuperscript{1}
\end{center}

\begin{center}
{\bf 1} Institute of Physics, University Greifswald, 
        17489 Greifswald, Germany
\\
{\bf 2} Department of Physics, Chiba University, Chiba 263-8522,
        Japan
\\
{\bf 3} The Rudolf Peierls Centre for Theoretical Physics,
        Oxford University, Oxford OX1 3NP, UK
\\
* ejima@physik.uni-greifswald.de
\end{center}

\begin{center}
 \today
\end{center}

\section*{Abstract}
{\bf We consider the dimerized spin-1 $XXZ$ chain with single-ion
anisotropy $D$. 
In absence of an explicit dimerization there are three
phases: a large-$D$, an antiferromagnetically ordered and a Haldane
phase. This phase structure persists up to a critical dimerization,
above which the Haldane phase disappears. We show that for weak 
dimerization the phases are separated by Gaussian and Ising quantum
phase transitions. One of the Ising transitions terminates in a
critical point in the universality class of the dilute Ising
model. We comment on the relevance of our results to experiments on
quasi-one-dimensional anisotropic spin-1 quantum magnets.
}

\vspace{10pt}
\noindent\rule{\textwidth}{1pt}
\tableofcontents\thispagestyle{fancy}
\noindent\rule{\textwidth}{1pt}
\vspace{10pt}

\section{Introduction}
It is well established that quantum effects in one-dimensional
antiferromagnetic (AFM) spin systems lead to interesting physical phenomena.
While a uniform Heisenberg chain is gapless for half-integer spins, an
exotic ground state with a finite gap appears for integer spins~\cite{Ha83}.  
For spins $S=1$, this Haldane phase can be understood 
in the framework of the Affleck-Kennedy-Lieb-Tasaki model~\cite{AKLT87,Affleck89}, 
whose exact ground state can be constructed in terms of 
valence bonds, i.e.,
singlet pairs of $S=1/2$ spins. 
Meanwhile, the Haldane phase is recognized as 
a symmetry-protected topological (SPT) state~\cite{GW09,PTBO10}
and attracts continued attention from both theoretical and experimental points
of view. For instance, the Haldane gap was confirmed experimentally 
in a compound with Ni$^{2+}$ ions 
Ni(C$_2$H$_8$N$_2$)$_2$NO$_2$(ClO$_4$)~\cite{PhysRevLett.56.371,RVRERS87}, in which 
a small value of the single-ion anisotropy $D$ was reported~\cite{DKLM91}.
A minimal model for the description of such anisotropic spin-1 chains is
\begin{eqnarray}
 \hat{H}_{XXZ,D}=
  J\sum_j (\hat{\bvec{S}}_j\cdot\hat{\bvec{S}}_{j+1})_\Delta
  +D\sum_j (\hat{S}_j^z)^2\, , 	  
 \label{xxz}
\end{eqnarray}
where 
$(\hat{\bvec{S}}_j\cdot\hat{\bvec{S}}_{j+1})_\Delta=
\hat{S}_j^x \hat{S}_{j+1}^{x} +\hat{S}_j^y \hat{S}_{j+1}^{y}
+\Delta \hat{S}_{j}^{z} \hat{S}_{j+1}^z$. 
Assuming a positive exchange parameter $J>0$ and $\Delta>0$, 
the ground-state phase diagram exhibits three gapped
phases~\cite{PhysRevB.67.104401}. At the isotropic point ($D=0$ and
$\Delta=1$) the model is in a Haldane phase. A sufficiently strong
single-ion anisotropy $D/J$ induces a Gaussian quantum phase
transition (QPT) with central charge $c=1$ to a topologically trivial
large-$D$ (LD) phase. On the other hand, increasing $\Delta$ for fixed
$D=0$ from the isotropic point leads to a Ising QPT with $c=1/2$ to a
long-range ordered AFM phase. At larger values of $\Delta$ and $D$
there is a first order transition between the LD and AFM phases.

A natural extension of the spin-1 $XXZ$ chain~\eqref{xxz} is the introduction of
an explicit bond alternation
\begin{eqnarray}
 \hat{H}=\hat{H}_{XXZ,D}+
  J\sum_j \delta(-1)^j(\hat{\bvec{S}}_j\cdot\hat{\bvec{S}}_{j+1})_\Delta\,.
 \label{model}
\end{eqnarray}
Interestingly this model realizes dimerized versions of the same three phases 
as the one described by Eq.~\eqref{xxz}, namely, dimerized Haldane
(D-H), AFM (D-AFM) and LD (D-LD) phases.  The case $D=0$ has been
studied previously~\cite{KNO96,KN97} and it was found that the
D-H to D-LD transition is again of the Gaussian type, but the {\it
  entire} D-AFM-phase boundary, including the transition to the D-LD
phase, belongs to the Ising universality class. A key question
is how the criticality at the phase boundary changes,  
if both $D$ and $\delta$ are finite. Earlier studies of half-filled
Hubbard-type models realizing SPT insulating and long-range ordered
(charge-density-wave) phases~\cite{LEF15,EELF16,ELEF17} indicated a
transition line that is separated into continuous Ising and
first-order QPTs. The meeting point of these lines belongs to the
tricritical Ising universality class with $c=7/10$, which can be
described by the second minimal model of conformal field
theory~\cite{FQS84,FQS85}.

In this paper, we determine and analyze the ground-state phase diagram of
the extended model~\eqref{model} by means of field theory and
matrix-product-state based density-matrix renormalization group
(DMRG)~\cite{White92,Sch11} techniques, focusing on the quantum
criticality at the phase boundaries.  
By calculating the central charge $c$, we provide compelling evidence
for the existence of a critical point in the tricritical Ising
universality class. Field-theory predictions for the phases and the
nature of the phase boundaries of the model~\eqref{model} with both
single-ion anisotropy $D$ and bond alternation $\delta$ are shown to
be in excellent agreement with numerical simulations. Finally, we
discuss the relevance of our results to experiments on dimerized
spin-1 materials~\cite{PhysRevLett.80.1312}.

\section{Ground-state phase diagram}
Let us first describe the numerical method we have used to determine the phase boundaries 
of the model~\eqref{model}. 
By means of the infinite DMRG (iDMRG)~\cite{Mc08} a characteristic correlation length $\xi_\chi$
can be calculated.
While this $\xi_\chi$ is always finite for fixed bond dimension $\chi$, 
it strongly peaks at a critical point and therefore allows for an accurate determination 
of QPT points, see Appendix~\ref{qpt-xi-cstar}.
This approach was already applied to half-filled Hubbard-type models~\cite{LEF15,EELF16,ELEF17}. 

In order to identify the different continuous phase transitions occurring in the model~\eqref{model}, 
we calculate the corresponding central charges $c$ via the entanglement entropy. 
For a critical system with $L$ sites and periodic boundary conditions, the von Neumann 
entanglement entropy of a contiguous block of $\ell$ sites with the rest of the system is
$ S_L(\ell)=\frac{c}{3}\ln
  \left[  \frac{L}{\pi}\sin\left(\frac{\pi\ell}{L}\right)
  \right]
  +s_1$,
where $s_1$ is a non-universal constant~\cite{CC04}. An accurate determination of 
the central charge is possible by using the relation~\cite{Ni11,EELF16}  
\begin{eqnarray}
 c^\ast(L) \equiv \frac{3[S_L(L/2-2)-S_L(L/2)]}{\ln\{\cos[\pi/(L/2)]\}}\;,
\label{cstar}
\end{eqnarray}
where in view of the explicit dimerization the doubled unit cell has been taken into account. 
Calculating the central charge numerically via Eq.~\eqref{cstar}, the universality classes of 
the QPT points are confirmed; this is demonstrated in Appendix~\ref{qpt-xi-cstar}.

For iDMRG simulations typical truncation errors are $10^{-12}$, using bond dimensions $\chi$
up to $1600$. In the case of finite-system DMRG calculations with periodic boundary conditions, 
e.g., by estimating the central charge via Eq.~\eqref{cstar}, the maximal truncation errors are 
about $10^{-9}$, with $\chi$ up to 6000.

\begin{figure}[tb]
 \centering
 \includegraphics[width=0.75\columnwidth,clip]{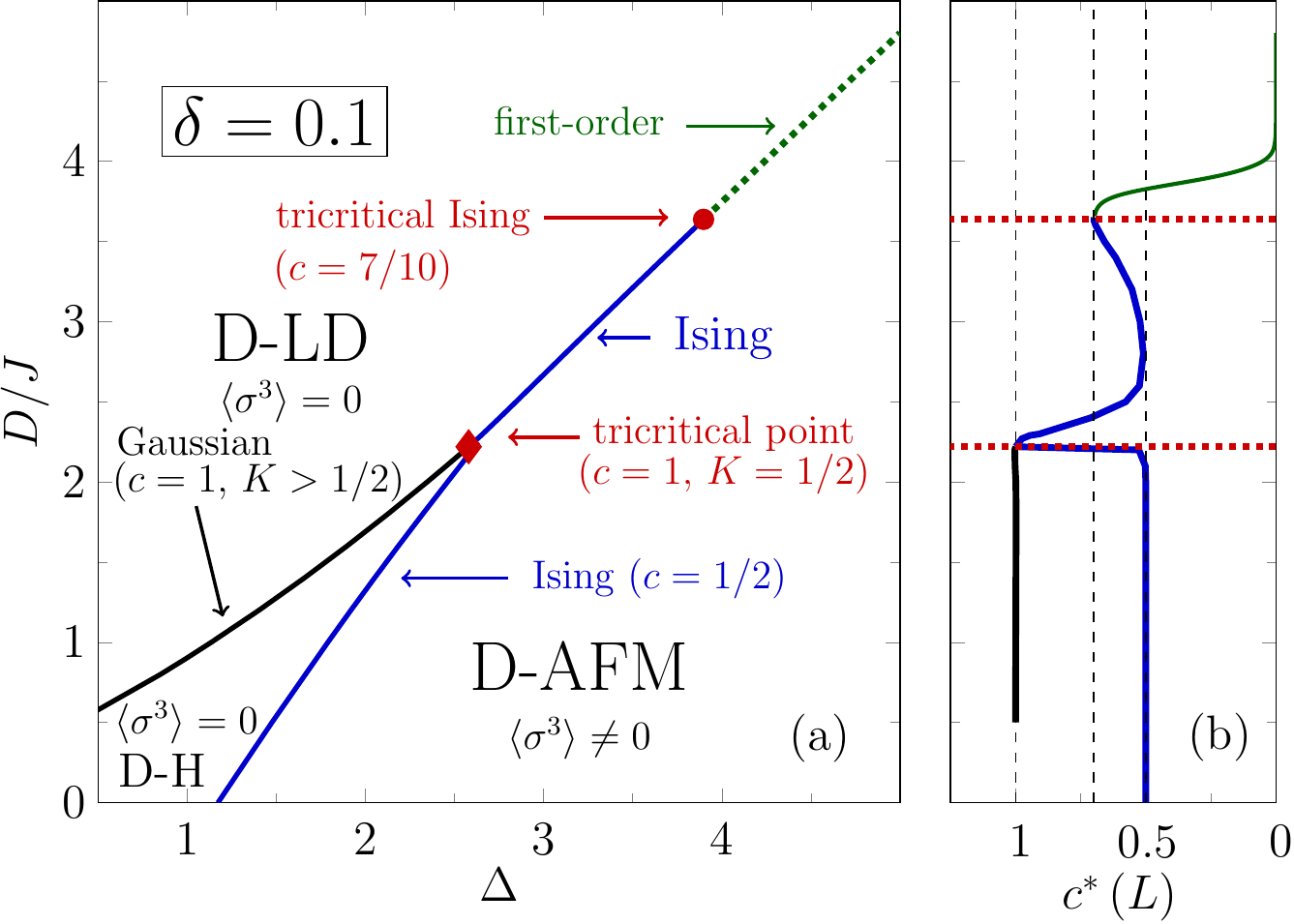}
 \caption{(a): Ground-state phase diagram of the model~\eqref{model} for $\delta=0.1$.
 The error bar of the tricritical (Ising) point is smaller than the symbol size. 
 $\langle\sigma^3\rangle$ denotes the third Ising order parameter, 
 determining the Ising QPT between the D-H or D-LD phase and the D-AFM phase.
 (b): Numerically obtained central charge $c^\ast(L)$ on various phase 
 transition lines from Eq.~\eqref{cstar} with $L=128$ and periodic boundary conditions.
 }
 \label{pd-delta0c1}
\end{figure}

Figure~\ref{pd-delta0c1}(a) shows the ground-state phase diagram of 
the model~\eqref{model} for $\delta=0.1$. 
For weak dimerization, the D-H phase survives between the D-LD and D-AFM phases. 
In contrast to the model without dimerization, however, the transition 
between the D-LD and D-AFM phases is continuous below a critical end point 
($\Delta_{\rm ce}$, $D_{\rm ce}/J$)$\simeq$(3.90, 3.64). 
Like the D-H{\trans}D-AFM line, this part of the transition belongs to 
the Ising universality class with central charge $c=1/2$, except for 
the critical end point, which belongs to the universality class 
of the tricritical Ising model with $c=7/10$. 
A tricritical Ising point at which the transition becomes first order 
is not observed in the dimerized model without single-ion anisotropy, 
simply because in this case the transition between the D-LD and D-AFM phases 
is always continuous. 
At the phase boundaries involving the Haldane phase, 
the universality classes are the same as in the non-dimerized model. 
Now the tricritical point, where the Haldane phase vanishes, 
is at 
($\Delta_{\rm tr}$, $D_{\rm tr}/J$)$\simeq$(2.58, 2.22). 
For $\delta \neq 0$, the central charge at this point is $c=1$. 

In the following, combining field theory and DMRG, we discuss various QPTs, 
including the direct Ising transition from the D-LD to the D-AFM phase.

\section{Field-theory approach}
In order to obtain a field-theory description of the model in the
vicinity of the various phase transition lines we consider the
Hamiltonian 
\begin{eqnarray}
 \hat{H}_{\rm FT}&=& \hat{H}
  -J\sum_j(1-\alpha)
  (\hat{\bvec{S}}_j\cdot\hat{\bvec{S}}_{j+1})^2_{\Delta^\prime}\,,
  \label{ft-hamil}
\end{eqnarray}
which differs from Eq.~\eqref{model} by an additional biquadratic exchange term. 
A field-theory description of the model~\eqref{ft-hamil} can be
constructed in the vicinity of the Takhtajan-Babujian
point\cite{Takhtajan82,Babujian82} ($\alpha=0$, $\delta=0$, $D=0$, $\Delta=1$ and
$\Delta^\prime=1$) following Ref.~\cite{Tsvelik90}. This leads
to a Hamiltonian density of the form 
\begin{eqnarray}
  \hat{{\cal H}}&=&\sum_{a=1}^3
   \frac{\mathrm{i}v_a}{2}
   [\hat{L}_a\partial_x \hat{L}_a-\hat{R}_a\partial_x \hat{R}_a]
   -\mathrm{i}m_a \hat{R}_a \hat{L}_a
    +\sum_{a=1}^3 g_a \hat{J}^a \hat{J}^a
   + \lambda \hat{\sigma}^1\hat{\sigma}^2\hat{\sigma}^3\, ,
   \label{ft-hamil-density}
\end{eqnarray}
where $\hat{L}_a$ and $\hat{R}_a$ are left and right moving Majorana fermions,
$\hat{\sigma}^a$ are three Ising order parameter fields and
\be
\hat{J}^a=-(\mathrm{i}/{2})\epsilon^{abc}[\hat{L}_b \hat{L}_c +
  \hat{R}_b \hat{R}_c].
\ee
The parameter $\lambda$ in $\hat{{\cal H}}$
is proportional to the dimerization $\delta$ and by virtue of the U(1)
symmetry of the microscopic Hamiltonian \eqref{ft-hamil} we have 
$v_1=v_2$, $m_1=m_2\equiv m$, and $g_1=g_2\equiv g$. The masses $m$
and $m_3$ are functions of $D$ and $\alpha$. The functional form of
this dependence is only known in the vicinity of the
Takhtajan--Babujian point and in what follows we therefore take $m_3$
and $m$ as free parameters, which we adjust in order to recover the
structure of the phase diagram obtained by DMRG. Our main working
assumption is that the field theory \fr{ft-hamil} remains a good
description of the low-energy degrees of freedom in the vicinity of
the various phase transition lines in the microscopic model even far
away in parameter space from the Takhtajan--Babujian point. We note
that an alternative way of deriving a field theory proposed by
Schulz\cite{Schulz86} leads to equivalent results. A
third approach would be to develop a field-theory description around
the SU(3) symmetric point of the spin-1 chain \cite{Affleck88,Itoi97,Assaraf99,Manmana11}, but we
do not pursue this here. The relation between lattice spin operators
and continuum fields is 
\be
\hat{S}_j^a\sim \hat{M}^a(x)+(-1)^j \hat{n}^a(x)\ ,
\ee
where $x=j a_0$ ($a_0$ is the lattice spacing). The smooth components
of the spin operators are proportional to the currents
$\hat{M}^a(x)\propto \hat{J}^a(x)$, while $\hat{n}^a(x)$ are expressed
in terms of Ising order and disorder operators as 
\begin{eqnarray}
 \hat{n}^x(x) &\propto& \hat{\sigma}^1(x)\hat{\mu}^2(x)\hat{\mu}^3(x)\,,
 \\ 
 \hat{n}^y(x) &\propto& \hat{\mu}^1(x)\hat{\sigma}^2(x)\hat{\mu}^3(x)\,,
 \\ 
 \hat{n}^z(x) &\propto& \hat{\mu}^1(x)\hat{\mu}^2(x)\hat{\sigma}^3(x)\,.
\end{eqnarray}
In order to facilitate comparisons between field-theory and iDMRG results for the
lattice model it is useful to define lattice operators
\begin{eqnarray}
 \hat{m}_j^{\alpha}=\frac{\hat{S}_j^\alpha+\hat{S}_{j+1}^\alpha}{2}\,,
  \ \ \
 \hat{n}_j^{\alpha}=(-1)^j\frac{\hat{S}_j^\alpha-\hat{S}_{j+1}^\alpha}{2}\,.
\end{eqnarray}  
At long distances we have
\be
\hat{m}_j^{\alpha}\approx\hat{M}^\alpha(x)\ ,\quad
\hat{n}_j^{\alpha}\approx\hat{n}^\alpha(x)\ .
\ee

It is convenient to use the U(1) symmetry to bosonize
\begin{eqnarray}
 \hat{L}_1+\mathrm{i}\hat{L}_2\sim
  \frac{1}{\sqrt{\pi a_0}}e^{-\mathrm{i}\sqrt{4\pi}\hat{\varphi}_L}\ ,
  \quad
 \hat{R}_1+\mathrm{i}\hat{R}_2\sim
  \frac{1}{\sqrt{\pi a_0}}e^{\mathrm{i}\sqrt{4\pi}\hat{\varphi}_R}\,.
\end{eqnarray}
In terms of the corresponding canonical Bose field
$\hat{\Phi}=\hat{\varphi}_L+\hat{\varphi}_R$ and the  
dual field $\hat{\Theta}=\hat{\varphi}_R-\hat{\varphi}_L$
the field theory \eqref{ft-hamil-density} reads:
\begin{eqnarray}
 \hat{{\cal H}}&=& {\cal \hat{H}}_3+{\cal \hat{H}}_B+{\cal \hat{H}}_{\textrm{int}}\,,
  \label{HBF}  
  \\ 
 {\cal \hat{H}}_3&=&\frac{\mathrm{i}v_3}{2}
  [\hat{L}_3\partial_x \hat{L}_3-\hat{R}_3\partial_x \hat{R}_3]
  -\mathrm{i}m_3\hat{R}_3\hat{L}_3\,,
  \\ 
 {\cal \hat{H}}_B&=&\frac{v}{2}\left[\frac{1}{K}(\partial_x\hat{\Phi})^2
			  +K(\partial_x\hat{\Theta})^2\right]
 -\frac{m}{\pi a_0}\cos\sqrt{4\pi}\hat{\Phi}\,,
  \\ 
 {\cal \hat{H}}_{\textrm{int}}&=&
  \frac{2\mathrm{i}g}{\pi a_0}\cos(\sqrt{4\pi}\hat{\Phi})\hat{L}_3\hat{R}_3
  +\lambda^\prime \sin(\sqrt{\pi}\hat{\Phi})\hat{\sigma}^3\,,
\end{eqnarray}
where $K$ is the Luttinger liquid (LL) parameter.
\subsection{Renormalization group analysis}
The most relevant perturbation is always the dimerization, and
concomitantly at weak coupling the $\lambda'$ term reaches strong
coupling first under the renormalization group (RG) flow. This results in a non-zero
dimerization  
\begin{equation}
  \langle\hat{d}\rangle\equiv\left\langle\frac{1}{L}\sum_j
  \hat{D}_j\right\rangle\neq 0
\ ,\quad \hat{D}_j=(-1)^j\bvec{\hat{S}}_j\cdot\bvec{\hat{S}}_{j+1}.
\end{equation}
For later convenience we define a lattice version of the
normal-ordered dimerization operator
\begin{equation}
  \hat{d}_j=\frac{\hat{D}_j+\hat{D}_{j+1}}{2}-\langle \hat{d}
  \rangle\ .
\end{equation}
To see what happens after the dimerization perturbation
has reached strong coupling we consider the next most relevant
operators, which are the Majorana mass term and the $\cos$-term in the
bosonic sector. Assuming that we have $m>0$, what happens then depends
on the sign of the Majorana mass term $m_3$. If it is positive the
third Ising model is in its disordered phase
$\langle\hat{\sigma}^3(x)\rangle=0$, while $m_3<0$ implies 
that $\langle\hat{\sigma}^3(x)\rangle\neq 0$. In the latter case the strong
coupling RG fixed point is amenable to a mean-field analysis. The term
$\hat{\cal H}_{\rm int}$ coupling the bosonic and fermionic sectors
can be decoupled, e.g.
\begin{eqnarray}
\hat{\sigma}^3(x)\sin\big(\sqrt{\pi}\Phi(x)\big)\rightarrow
\langle\hat{\sigma}^3(x)\rangle\sin\big(\sqrt{\pi}\hat{\Phi}(x)\big)
+\hat{\sigma}^3(x)\langle\sin\big(\sqrt{\pi}\hat{\Phi}(x)\big)\rangle\ .
\end{eqnarray}
This leads to a mean-field description in terms of an Ising model in a
longitudinal field and 
a double sine-Gordon model~\cite{FABRIZIO2000647,DELFINO1998675}
\bea
\hat{\cal H}_{\rm MF}&=&\frac{\mathrm{i}v_3}{2}
  [\hat{L}_3\partial_x \hat{L}_3-\hat{R}_3\partial_x \hat{R}_3]
  -\mathrm{i}\widetilde{m}_3\hat{R}_3\hat{L}_3+h\hat{\sigma}^3
+\frac{v}{2}\left[\frac{1}{K}(\partial_x\hat{\Phi})^2
			  +K(\partial_x\hat{\Theta})^2\right]
 \nn
 &&-\frac{\widetilde{m}}{\pi a_0}\cos(\sqrt{4\pi}\hat{\Phi})
 +\widetilde{\lambda}\sin(\sqrt{\pi}\hat{\Phi})\ ,
 \eea
 where
 \bea
 \widetilde{\lambda}&=&\lambda'\langle\hat{\sigma}^3\rangle\ ,\quad
 h=\lambda'\langle\cos(\sqrt{4\pi}\hat{\Phi})\rangle\ ,\nn
\widetilde{m}&=&m+2ig\langle \hat{R}_3\hat{L}_3\rangle\ ,\quad
\widetilde{m_3}=m_3+\frac{2g}{\pi a_0}\langle\cos(\sqrt{4\pi}\hat{\Phi})\rangle .\quad
 \eea
The classical ground state of the double sine-Gordon model is obtained
by solving 
\be
\frac{2\widetilde{m}}{\pi}\sin(\sqrt{4\pi}\hat{\Phi}_c)
+\widetilde{\lambda}\cos(\sqrt{\pi}\hat{\Phi}_c)= 0\ .
\ee
Importantly, this tells us that for $\widetilde{m}>0$ we have
\be
\langle\cos(\sqrt{\pi}\hat{\Phi}(x))\rangle\neq 0\ ,
\ee
which in turn implies that
\be
\langle \hat{n}^z(x)\rangle\propto\langle
\hat{\sigma}^3(x)\cos(\sqrt{\pi}\hat{\Phi})\rangle\neq 0\ .
\ee
Hence the strong coupling RG fixed point describes a phase where
antiferromagnetic order coexists with dimerization. This is the D-AFM
phase identified above by the DMRG. 

In the other phases the RG fixed points again occur at strong coupling
but cannot be analyzed in terms of a simple mean-field
argument. However, the field theory nevertheless allows for a
description of the various transition lines as shown in what follows.
\subsection{Quantum phase transitions}

\subsubsection{D-LD \texorpdfstring{\trans}{\trans} D-AFM phase transition line}

This corresponds to the
situation where the bosonic sector remains gapped, while the third
Ising model undergoes a transition between a disordered phase
$\langle\hat{\sigma}^3\rangle=0$ on the D-LD side and an ordered phase
$\langle\hat{\sigma}^3\rangle\neq 0$ on the D-AFM side of the phase
diagram. As a result the D-LD{\trans}D-AFM phase transition is in
the universality class of the two-dimensional Ising model. In the
vicinity of the transition we may project onto the low-energy Ising
degrees of freedom following e.g. Ref.\cite{Wang02}. Details are
given in Appendix \ref{app:lowE}. This yields
\begin{eqnarray}
\hat{m}_j^z\Bigl|_{\rm low}&=&A\partial_x\hat{\sigma}^3(x)+\dots\ ,\\ 
\hat{n}_j^z\Bigl|_{\rm low}&=&B\hat{\sigma}^3(x)+\dots\ , \\ 
\hat{d}_j\Bigl|_{\rm low}&=&\mathrm{i}C\hat{R}_3(x)\hat{L}_3(x)+\dots\ .
\end{eqnarray}
Along the phase transition line we thus have
\begin{eqnarray}
\langle \hat{n}_{j}^z \hat{n}_{j+\ell}^z\rangle &=& 
B^2{\ell}^{-1/4}+\dots\ ,
\label{njz-njlz} \\ 
\langle \hat{m}_{j}^z \hat{n}_{j+\ell}^z\rangle &=& 
-\frac{AB}{4}\ell^{-5/4}+\dots\ ,
\label{mjz-mjlz}\\ 
\langle \hat{m}_{j}^z \hat{m}_{j+\ell}^z\rangle &=&
\frac{5A^2}{16}\ell^{-9/4}+\dots\ , 
\label{SzSz}
\end{eqnarray}
and
\be
\langle\hat{d}_j\hat{d}_{j+\ell}\rangle=C^2\ell^{-2}+\ldots\,. 
\label{dimer-dimer}
\ee
The predictions~\eqref{njz-njlz}--\eqref{dimer-dimer} are compared to
iDMRG simulations below.

\subsubsection{D-H \texorpdfstring{\trans}{\trans} D-AFM phase transition line}
The D-AFM to D-H transition is described by the same scenario as
discussed above, since it also belongs to the Ising universality class with $c=1/2$. 
Accordingly, Eqs.~\eqref{njz-njlz}--\eqref{dimer-dimer} are valid on
this transition line as well.

\subsubsection{D-H \texorpdfstring{\trans}{\trans} D-LD phase transition line}
As we cross from the D-AFM into the D-H phase at fixed $\Delta$ by
increasing $D$ the (effective) Majorana mass $m_3$ increases. Assuming
that this relation continues to hold, the characteristic energy scale
in the Majorana sector can eventually become large compared to that of
the bosonic sector and it is then justified to integrate out the
Majorana sector. This leads to an effective low-energy description
in terms of a sine-Gordon model
\be
\hat{\cal H}_{\rm low}=
\frac{v}{2}\left[\frac{1}{K}(\partial_x\hat{\Phi})^2
			  +K(\partial_x\hat{\Theta})^2\right]
 -\frac{m^*}{\pi a_0}\cos(\sqrt{4\pi}\hat{\Phi})\ .
\ee
The main effect of integrating out the Majorana sector is the
renormalization of the sine-Gordon coupling. Importantly, $m^*$
can vanish for particular values of $D$, which corresponds to a phase
transition line described by a LL characterized by the LL parameter $K$. 
The low-energy projections of the lattice spin operators along this line are
\begin{eqnarray}
\hat{d}_j\Bigl|_{\rm low}&=& A_D\cos\big(\sqrt{4\pi}\hat{\Phi}(x)\big)+\dots\ ,\\ 
\hat{n}_j^z\Bigl|_{\rm low}&=&
A_z\ \sin\left(\sqrt{4\pi}\hat{\Phi}(x)\right)+\dots\ ,  \label{nxynxy}\\
\hat{n}_j^x\Bigl|_{\rm low}&=&A_x\
\cos\left(\sqrt{\pi}\hat{\Theta}(x)\right)+\dots\ ,\\ 
\big(S^+_j\big)^2\Bigl|_{\rm low}&=& A_2\
e^{\mathrm{i}\sqrt{4\pi}\hat{\Theta}(x)}+\dots\ ,\\ 
\hat{m}_j^x\Bigl|_{\rm low}&=&\frac{a_0}{\sqrt{\pi}}\partial_x\hat{\Phi}(x)+\dots\,.
\end{eqnarray}
This gives the following field-theory predictions for power-law decays
of two-point functions 
\begin{eqnarray}
 \langle \hat{n}_j^z\hat{n}_{j+\ell}^z\rangle&=& \frac{A_z^2}{2}\ell^{-2K}+\dots\,,
  \label{nznz}\\
 \langle \hat{n}_j^\alpha\hat{n}_{j+\ell}^\alpha\rangle&=&\frac{A_x^2}{2} \ell^{-1/2K}+\dots\,,
 \ \ \
 \alpha=x,\,y\,,\\
 \langle (\hat{S}_j^+)^2
 (\hat{S}_{j+\ell}^-)^2\rangle&=&A_2^2\ell^{-2/K}+\dots\,,\label{splussminus}\\
 \langle \hat{m}_j^z\hat{m}_{j+\ell}^z\rangle&=& \frac{K}{2\pi^2}\ell^{-2}+\dots\,,
 \label{sq-trans}\\
 \langle\hat{d}_j\hat{d}_{j+\ell}\rangle&=& \frac{A_D^2}{2}\ell^{-2K}+\dots\,.
 \label{dd}
\end{eqnarray}
      
\section{DMRG analysis}
In this section, we examine various two-point correlation 
functions of the lattice Hamiltonian~\eqref{model} using iDMRG 
in order to prove the field-theory predictions described in the last section. 
Then, the topological properties of each phase are discussed by simulating
topological order parameters. 

\subsection{Quantum phase transitions}

\begin{figure}[ht]
 \centering
 \includegraphics[width=0.95\columnwidth,clip]{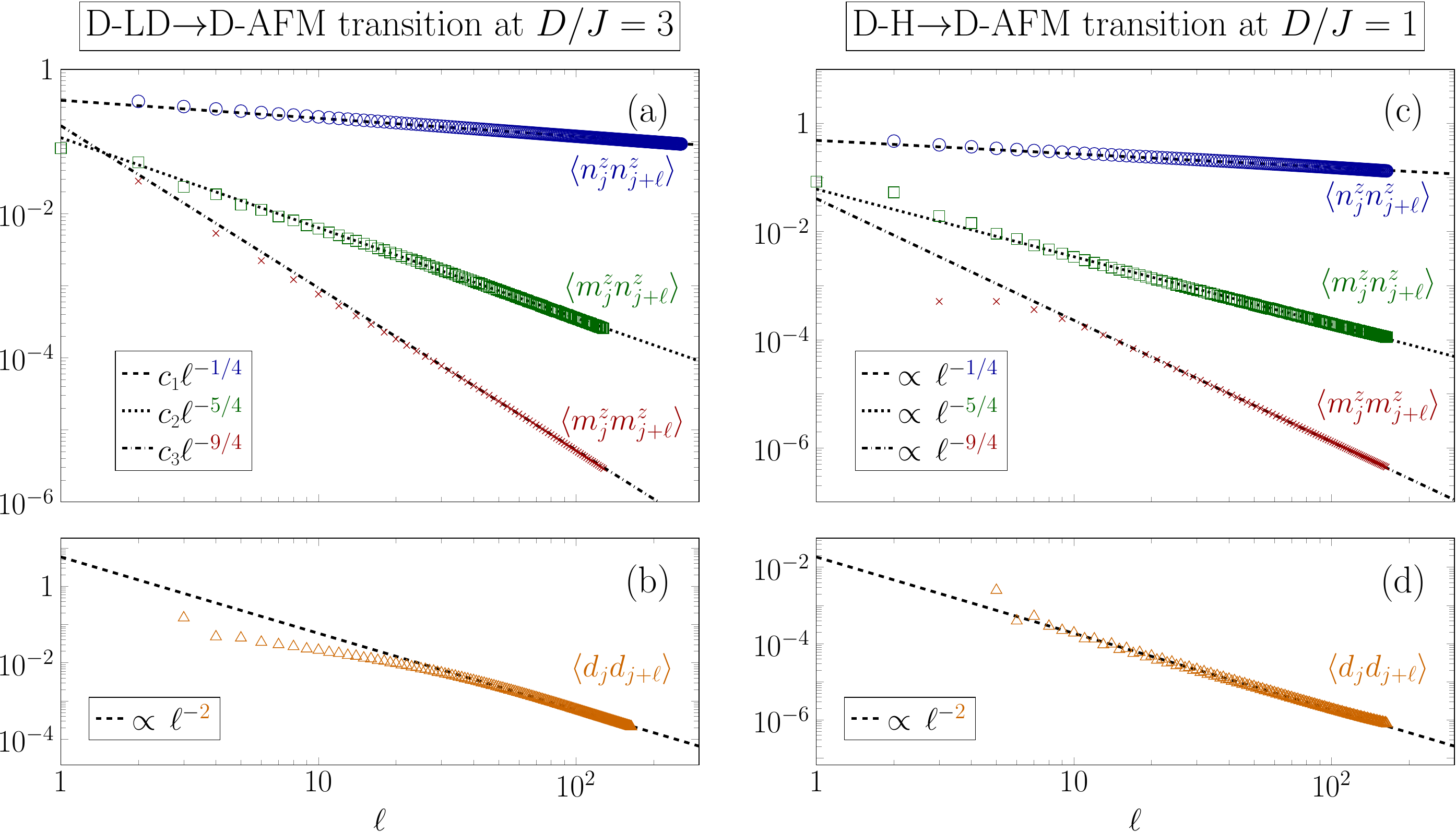}
 \caption{The connected longitudinal spin-spin (upper panels) and 
 dimerization (lower panels) two-point functions at the Ising transition 
 point for fixed $D/J=3$ (left panels) and $D/J=1$ (right panels) 
 with $\delta=0.1$, obtained by iDMRG with $\chi=1600$. 
 Correlation functions (symbols) show a power-law decay 
 in accordance with the field-theory predictions
 Eqs.~\eqref{njz-njlz}--\eqref{dimer-dimer} [lines].
 }
 \label{dd-szsz}
\end{figure}
\subsubsection{D-LD \texorpdfstring{\trans}{\trans} D-AFM and 
D-H \texorpdfstring{\trans}{\trans} D-AFM Ising phase transition lines}
%
For fixed $D/J=3$ and $\delta=0.1$ the Ising QPT
occurs at $\Delta_{\rm c}\simeq3.303$ between D-LD and D-AFM phases
as extracted from correlation length $\xi_\chi$. 
At this transition point various two-point functions can be computed
by iDMRG. Here, $\chi=1600$. As shown in Fig.~\ref{dd-szsz}(a)
field-theory predictions for diverse two-point functions 
of $z$-component spin operators~~\eqref{njz-njlz}--\eqref{SzSz} can be proved by iDMRG.
Figure~\ref{dd-szsz}(b) demonstrates that also 
the dimer-dimer correlation function is in agreement with the
power-law behavior according to Eq.~\eqref{dimer-dimer} for large
distances $\ell\gg 1$. 

The relations between the coefficients in
Eqs.~~\eqref{njz-njlz}--\eqref{dimer-dimer} can be verified by
fitting the iDMRG data to the field-theory predictions.
For instance, in the case of the D-LD{\trans}D-AFM transition 
at $D/J=3$ [Fig.~\ref{dd-szsz}(a)], we obtain $c_1\simeq 0.381$ ($B\simeq0.617$) 
and $c_3\simeq0.158$ ($A\simeq0.711$), 
i.e., $AB/4\simeq0.110$, which is in good agreement with $c_2\simeq0.114$ 
from Eq.~\eqref{mjz-mjlz}.

Along the Ising critical line separating the D-H and D-AFM phases the
long-distance behavior of these correlation functions determined by
iDMRG is again in
excellent agreement with field-theory predictions, \emph{cf.}
Eqs.~\eqref{njz-njlz}-\eqref{dimer-dimer}. A representative example is
shown in Figs.~\ref{dd-szsz}(c) and (d) for $D/J=1$ and $\Delta_{\rm
  c}\simeq1.789$. 

\subsubsection{D-H \texorpdfstring{\trans}{\trans} D-LD phase transition line}
\label{DH-DAFM}
\begin{figure}[htb]
 \centering
 \includegraphics[width=0.95\columnwidth,clip]{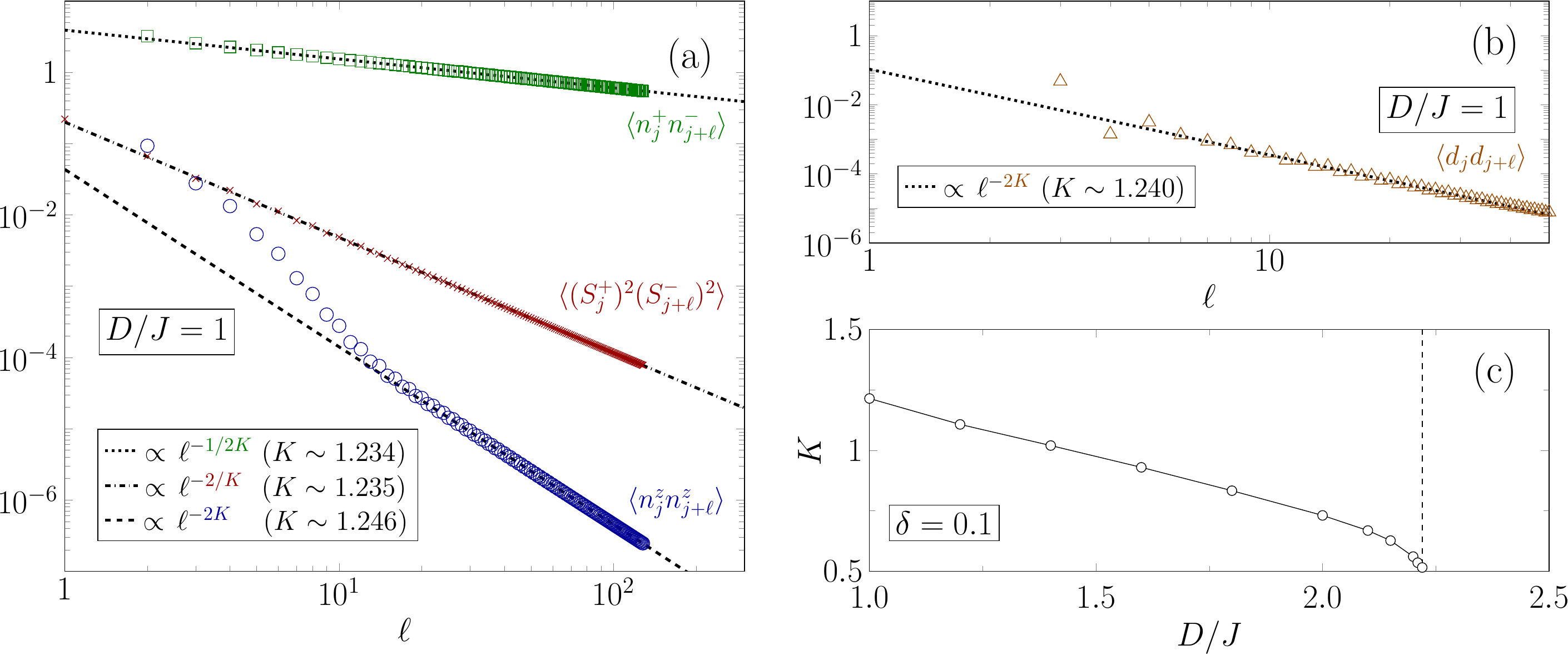}
 \caption{
 Spin-spin (a) and dimer-dimer (b) correlation functions 
 at the $c=1$ transition 
 for $D/J=1$ and $\delta=0.1$ computed by iDMRG with bond dimension $\chi=1600$.
 The extracted values of the LL parameter $K$ are in good agreement. 
 (c) Extrapolated values of LL parameters $K$
 via $S(q)$ of Eq.~\eqref{Sq} on the $c=1$ transition line 
 for $\delta=0.1$, obtained by DMRG with up to $L=1024$ sites
 and open boundary conditions.
 }
 \label{SPT-D-corr-ksigma}
\end{figure}

Along the line of Gaussian QPTs separating the D-H and D-LD phases
the exponents characterizing the long-distance behavior of
correlation functions depends on the LL parameter $K$ as described in
Eqs.~\eqref{nznz}-\eqref{splussminus} and \eqref{dd}. 
In order to facilitate a comparison to the field-theory results 
we therefore require the LL parameter $K$. For fixed $D/J=1$ the
Gaussian transition occurs at $\Delta_{\rm c}\simeq 1.135$. In
Figs.~\ref{SPT-D-corr-ksigma}(a) and (b) we show numerical results of
correlation functions obtained by iDMRG. The values of LL parameters
extracted from the fits to Eqs.~\eqref{nznz}-\eqref{splussminus} 
and \eqref{dd} show reasonable agreement with each other. 

These values can also be extracted from the long-distance behavior 
of the smooth part of the spin-spin correlation function~\eqref{sq-trans}, 
that is, the LL parameter determines the amplitude of 
the correlation function but not the exponent. 
We calculate the longitudinal spin correlation function 
and isolate the smooth component from a Fourier transformed structure factor
\begin{eqnarray}
 S(q)=\frac{1}{L} 
  \sum_{j\ell}e^{\mathrm{i}q(j-\ell)}
  \left(\left\langle \hat{S}^z_j \hat{S}^z_{\ell} \right\rangle
   -\left\langle \hat{S}^z_j \right\rangle \left\langle \hat{S}^z_\ell 
     \right\rangle\right) \; 
  \label{Sq}
\end{eqnarray}
for $q\approx 0$, where $q=2\pi/L$.
The LL parameter is determined as $K=\lim_{q \to 0}\pi S(q)/q$~\cite{EGN05}. 
Figure~\ref{SPT-D-corr-ksigma}(c) shows the results for the Luttinger
parameter $K$ on the $c=1$ line for $\delta=0.1$. 
At $\Delta=1$ we have $K=1.215$, in reasonable agreement with
the values obtained from the exponents of correlation functions 
in Figs.~\ref{SPT-D-corr-ksigma}(a) and (b). Following the Gaussian transition
line by increasing $\Delta$ and $D/J$ the Luttinger parameter
decreases and takes the value $K\simeq 1/2$ at the point when the
Gaussian line merges with the line of Ising QPTs.

\subsection{Topological order parameters}
Let us now explore the topological properties of the phases  
of the model~\eqref{model}.
Following Vidal~\cite{PhysRevLett.98.070201}, we use the 
infinite matrix-product-state representation formed by $\chi\times\chi$
matrices $\Gamma_\sigma$ and a positive real, diagonal matrix $\Lambda$:
\begin{eqnarray}
 |\psi\rangle=\sum_{\cdots\sigma_j,\sigma_{j+1}\cdots}
  \cdots\Lambda\Gamma_{\sigma_j}\Lambda\Gamma_{\sigma_{j+1}}\cdots
  |\cdots\sigma_j,\sigma_{j+1},\cdots\rangle\,,
\end{eqnarray}  
where the index $\sigma$ labels the basis states of the local Hilbert
spaces. The $\Gamma_{\sigma}$ and $\Lambda$ are assumed to be 
in the canonical form: 
\begin{eqnarray}
 \sum_\sigma \Gamma_\sigma \Lambda^2 \Gamma_\sigma^\dagger
  =\mathds{1}
  =\sum_\sigma \Gamma_\sigma^\dagger \Lambda^2 \Gamma_\sigma\,.
\end{eqnarray}
If $|\psi\rangle$ is invariant under an internal symmetry represented 
by a unitary matrix $\Sigma_{\sigma\sigma^\prime}$, then the transformed
$\Gamma_\sigma$ matrices satisfy~\cite{PhysRevLett.100.167202,PTBO10} 
\begin{eqnarray}
 \sum_{\sigma^\prime}\Sigma_{\sigma\sigma^\prime}\Gamma_{\sigma^\prime}
  =e^{\mathrm{i}\theta}U^\dagger\Gamma_{\sigma}U\,.
\end{eqnarray}
Here $U$ is a unitary matrix that commutes with $\Lambda$,
and $e^{\mathrm{i}\theta}$ is a phase factor. 
In the case of time reversal symmetry (inversion symmetry), $\Gamma_\sigma$
on the left-hand side is replaced by its complex conjugate 
$\Gamma_\sigma^{\dagger}$ (its transpose $\Gamma_\sigma^{T}$). 
Exploiting the properties of the matrices $U$ each SPT phase 
can be classified~\cite{PTBO10}: In the case of time reversal (inverse)
symmetry the matrices satisfy 
$U_{\cal T}^{\phantom{\cal T}}U_{\cal T}^\ast=\pm\mathds{1}$
($U_{\cal I}^{\phantom{\cal I}}U_{\cal I}^\ast=\pm\mathds{1}$), 
and the sign can be used to distinguish different SPT phases.
In presence of a 
$\mathds{Z}_2\times\mathds{Z}_2$ symmetry the order parameter
is given by 
\begin{eqnarray}
  O_{\mathds{Z}_2\times\mathds{Z}_2}
   =\frac{1}{\chi}\mathrm{Tr}
     \left(U_x^{\phantom{\dagger}}U_z^{\phantom{\dagger}}U_x^{\dagger}U_z^{\dagger}\right)\, ,
\end{eqnarray}
where we use the symmetry operations 
$\hat{R}^x=\exp(\mathrm{i}\pi\sum_j\hat{S}_j^x)$ and 
$\hat{R}^z=\exp(\mathrm{i}\pi\sum_j\hat{S}_j^z)$
to calculate $U_x$ and $U_z$.

\begin{figure}[htb]
 \centering
 \includegraphics[width=0.6\columnwidth,clip]{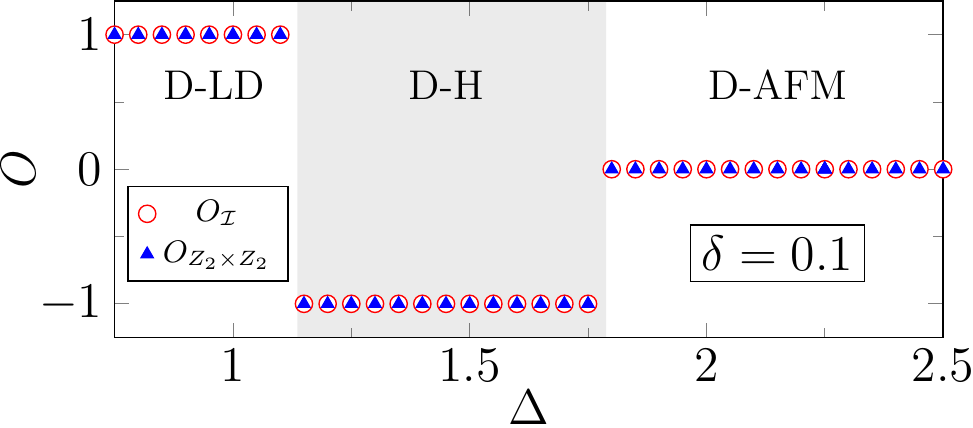}
 \caption{Topological order parameters for inversion symmetry $O_{\cal I}$
 and $\mathds{Z}_2\times\mathds{Z}_2$ spin rotation symmetry 
 $O_{\mathds{Z}_2\times\mathds{Z}_2}$ at $D/J=1$ and $\delta=0.1$.
 }
 \label{topo-order-para}
\end{figure}

In the presence of dimerization the unit cell consists of two sites, which we have to 
block together in order to apply the above description. For the model~\eqref{model}, 
blocking sites across weak bonds gives the same values of the order parameters as blocking 
across strong bonds. Figure~\ref{topo-order-para} shows the iDMRG results for the
order parameters in case of inverse and $\mathds{Z}_2\times\mathds{Z}_2$
symmetries. If $U_{x}$ and $U_{z}$ commute 
($O_{\mathds{Z}_2\times\mathds{Z}_2}=1$), the system is 
in a trivial phase, i.e., a site-factorizable LD state, whereas 
if they anticommute ($O_{\mathds{Z}_2\times\mathds{Z}_2}=-1$), 
the system realizes a non-trivial Haldane state. 
If the symmetry is broken, we set $O_{\mathds{Z}_2\times\mathds{Z}_2}=0$. 
Obviously, the order parameter $O_{\mathds{Z}_2\times\mathds{Z}_2}$ changes its sign
only if a phase transition occurs between D-LD and D-H phases.
$O_{\cal I}$ behaves similarly to $O_{\mathds{Z}_2\times\mathds{Z}_2}$,
i.e., $O_{\cal I}=\pm 1$ for the two symmetric
phases, and $O_{\cal I}=0$ in the D-AFM phase.

To summarize this subsection, dimerization does not affect
the topological properties of the system~\eqref{model},
so that the D-H (D-LD) phase remains a non-trivial (trivial)
SPT phase as in the system without dimerization~\eqref{xxz}. 

\section{Relevance to experiments}
Let us finally relate our findings with experimental results. 
There are several realizations of spin-1 bond-alternating chains, 
such as Ni(C$_9$H$_24$N$_4$) (NO$_2$)ClO$_4$~\cite{NTENP,NTENP2} 
and {[}Ni(333-tet)($\mu$-N$_3$)$_n${]}(ClO$_4$)$_n$~\cite{VERS92,EVRFSF94,EVSF94}.
Most remarkably, in the latter material a logarithmic decrease 
of the susceptibility was observed at low temperature, 
indicating a vanishing excitation gap~\cite{PhysRevLett.80.1312}. 
Comparing quantum Monte-Carlo simulations with experimental data
suggested that the material is described by a Hamiltonian of the form
(\ref{model}) with $\delta=0.25$, $\Delta=1$ and $D/J=0$.
Totsuka {\it et al.}~\cite{TNHS95} determined the critical point for $D=0$
numerically and obtained $\delta_{\rm c}=0.25\pm 0.01$ and $c=1$,
while results by Kitazawa and Nomura~\cite{KN97} suggested that $\delta_{\rm c}=0.2598$.  
Importantly these parameter sets are close to the location of
the point where the Gaussian and Ising phase transitions merge~\cite{KNO96,KN97}. 

\begin{figure}[ht]
 \centering
 \includegraphics[width=0.95\columnwidth,clip]{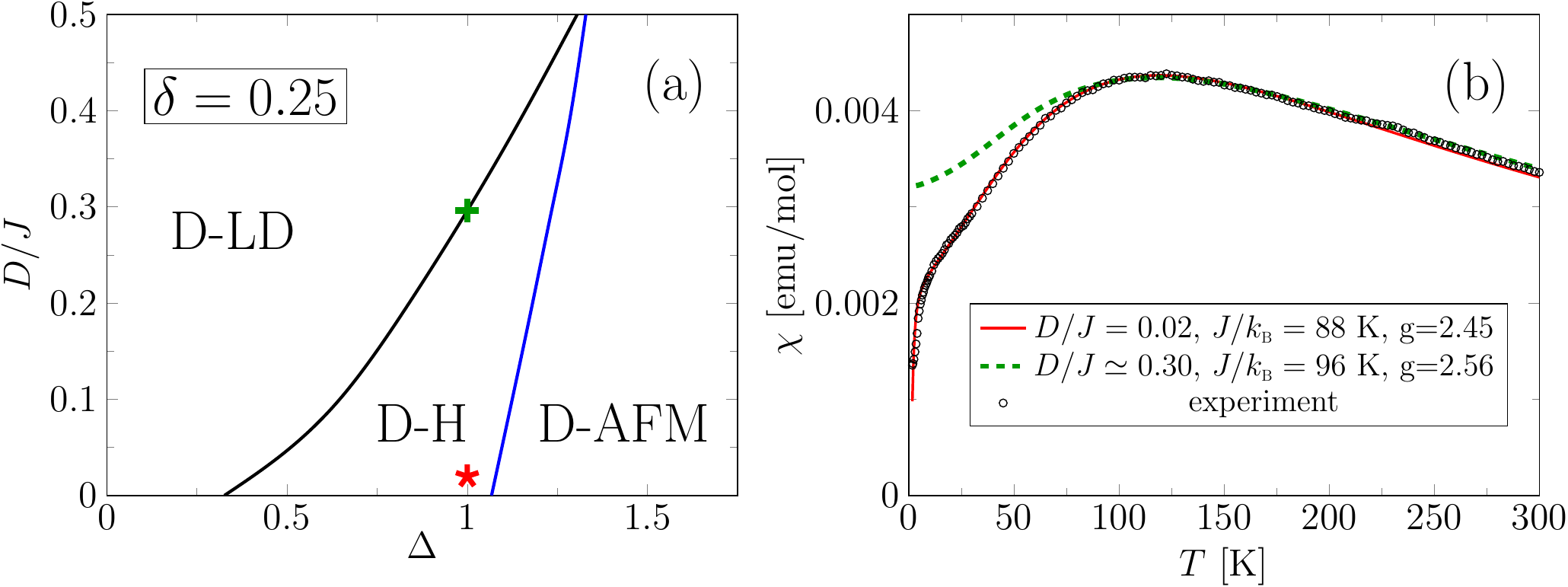}
 \caption{(a) Ground-state phase diagram of the model~\eqref{model}
 for $\delta=0.25$. The red star denotes the parameter set 
 corresponding to the Ni compound  
 {[}Ni(333-tet)($\mu$-N$_3$)$_n${]}(ClO$_4$)$_n$, and the green cross
 gives the Gaussian transition point 
 [$(D/J)_{\rm c}\simeq 0.296$] 
 for fixed $\Delta=1$. (b) Temperature dependence of the magnetic
 susceptibility of the powdered sample for 
 {[}Ni(333-tet)($\mu$-N$_3$)$_n${]}(ClO$_4$)$_n$ (circles)
 taken from Ref.~\cite{PhysRevLett.80.1312}.
 The red solid line is the iTEBD data for $\Delta=1$, $D/J=0.02$ and 
 $\delta=0.25$ with $J/k_{\rm B}=88$ K and $g=2.45$.
 For comparison, we also show the iTEBD result at the Gaussian
 transition for fixed $\Delta=1$ (green dashed line).  
 }
 \label{delta0c25}
\end{figure}

In the following, we therefore determine the ground-state phase diagram 
of the model~\eqref{model} for $\delta=0.25$ and reexamine the magnetic susceptibility of 
the above mentioned nickel compound using the infinite time-evolving block decimation 
(iTEBD)~\cite{PhysRevLett.98.070201}.
Figure~\ref{delta0c25}(a) displays the corresponding phase diagram of the 
model~\eqref{model}. Although the extent of the Haldane phase is significantly
reduced, the Gaussian and Ising transition lines can still be detected numerically.
As shown in Fig.~\ref{delta0c25}(b) the experimental data of the magnetic susceptibility 
for {[}Ni(333-tet)($\mu$-N$_3$)$_n${]}(ClO$_4$)$_n$
can be fitted most successfully for $\Delta=1$ and $D/J=0.02$, taking the reported small 
single-ion anisotropy $D/J<0.1$~\cite{PhysRevLett.80.1312} into account.
On the other hand, the numerical data at the Gaussian transition point
for fixed $\Delta=1$ deviates from experimental ones in the
lower-temperature regime. Thus, this nickel compound may be even
closer to the Ising transition line than to the $c=1$ transition line
considered so far. 
It would be interesting to investigate signatures of the  Ising QPT experimentally, 
e.g., by inelastic neutron scattering, where the corresponding dynamical structure 
factor can be calculated numerically, see Ref.~\cite{EF15}.

\section{Summary and Conclusions}
In this work we investigated the ground-state phase diagram and 
quantum criticality of the dimerized spin-1 $XXZ$ chain with single-ion 
anisotropy $D$, employing a combination of analytical 
and numerical techniques. 
For weak dimerization ($\delta\lesssim 0.26$) and single-ion anisotropy, 
the symmetry-protected topological Haldane phase survives and 
the transition between the D-LD and D-AFM phases, 
which is always of first order in the absence of dimerization, 
becomes partially continuous. 
The continuous section of the transition line belongs to the Ising universality class 
with central charge $c=1/2$. With increasing the magnitude of $D$, 
this Ising line terminates at a tricritical Ising point with $c=7/10$, 
above which the phase transition becomes first order.
A comprehensive description of the phases and phase boundaries can be achieved 
by a bosonization-based field theory including three Majorana fermions. 
The field-theory predictions for various correlation functions have been confirmed 
by numerical iDMRG calculations. 

Finally, we have revisited the experimental results for 
the Ni compound {[}Ni(333-tet)($\mu$-N$_3$)$_n${]}(ClO$_4$)$_n$
showing gapless behavior and have demonstrated that 
the corresponding parameter set might be not only 
in the vicinity of the Gaussian transition line but also
very close to the Ising transition line. 
Further experimental research for this material, such as neutron scattering, 
would be desirable.

\section*{Acknowledgements}
We thank M. Hagiwara for useful discussions and providing us 
with their experimental data.
The iDMRG simulations were performed using the ITensor library~\cite{ITensor}.

\paragraph{Funding information}
This work was supported by the Deutsche Forschungsgemeinschaft
(Germany) under Grant No. FE 398/8-1 (FL), by the EPSRC under Grant No.
EP/N01930X (FHLE) and the National Science Foundation under Grant
No. NSF PHY-1748958 (FHLE). 
FHLE is grateful to the Erwin Schr\"odinger International Institute 
for Mathematics and Physics for hospitality and support 
during the programme on \emph{Quantum Paths}. 
TY acknowledges support by a Chiba University SEEDS Fund 
and YO acknowledges support by a Grant-in-Aid for Scientific Research 
(Grant No. 17K05530) from JSPS of Japan.
HF is grateful to the Los Alamos National Laboratory for hospitality 
and support.

\appendix
\section{Low-energy projections of operators}
\label{app:lowE}
Let us denote the Euclidean action corresponding to the Hamiltonian
density \fr{HBF} by
\be
S=S_3+S_B+S_{\rm int}\ ,
\ee
where $S_3$ and $S_B$ involve only Ising and bosonic degrees of
freedom respectively and $S_{\rm int}$ describes the interaction
between the two sectors. 
In the regimes where the mass scale
associated with $S_3$ is much smaller (larger) than the one associated
with $S_B$ and where $S_{\rm int}$ can be treated as a perturbation, 
we may integrate out the bosonic (fermionic) degrees of freedom,
see e.g. Ref.~\cite{Wang02}.

\subsection{Integrating out the bosonic degrees of freedom}
This case pertains to the transition lines between the D-AFM phase and
the D-LD and D-H phases. In these cases the low-energy projection
of a general local operator is given by 
\be
\hat{O}\Bigl|_{\rm low}=\int {\cal D}\Phi\ e^{-S_B}e^{-S_{\rm
    int}}\hat{O}=\langle{\hat{O}}\rangle_{\Phi}
-\langle S_{\rm int}\hat{O}\rangle_{\Phi}+\dots\ ,
\ee
where $\langle\rangle_{\Phi}$ denotes the average with respect to the
bosonic action $S_B$. As we have assumed that the parameter $m$ is
positive, we have
\be
\langle\sin(\sqrt{4\pi}\Phi)\rangle_{\Phi}=0.
\ee
This implies that the low-energy projection of the dimerization operator is
\bea
\hat{D}_j\Bigl|_{\rm low}&\sim&
-\langle S_{\rm
  int}\hat{\sigma}^3(x)\sin\big(\sqrt{\pi}\hat{\Phi}(x)\big)\rangle_{\Phi}+\dots\ \nn
&=&-\lambda'\int d\tau dy\ \hat{\sigma}^3(x)\hat{\sigma}^3(y,\tau) 
\langle\sin\big(\sqrt{\pi}\hat{\Phi}(x,0)\big)
\sin\big(\sqrt{\pi}\hat{\Phi}(y,\tau)\big)\rangle_{\Phi}+\dots\nn
&=& \langle \hat{d}\rangle+i C \hat{R}_3(x)\hat{L}_3(x)+\dots\,.
\eea
In the last line we have used that the expectation value in the bosonic sector
decays exponentially in the Euclidean distance $r=\sqrt{(x-y)^2+v^2\tau^2}$\,,
\be
\langle\sin\big(\sqrt{\pi}\hat{\Phi}(x,0)\big)
\sin\big(\sqrt{\pi}\hat{\Phi}(y,\tau)\big)\rangle_{\Phi}\propto
e^{-r/\xi}\ ,
\ee
which in turn allows us to employ the operator product expansion in the Ising sector
\be
\hat{\sigma}^3(x)\hat{\sigma}^3(y,\tau)=
\left(\frac{a_0}{r}\right)^\frac{1}{4}\left[1-\mathrm{i}\pi r \hat{R}_3(x)\hat{L}_3(x)\right]+\dots\,.
\ee
Finally we have fixed the constant part in the low-energy projection
by using that it must give the correct expectation value of the
dimerization operator. Similarly we obtain
\bea
\hat{M}^z_j\Bigl|_{\rm low}&\sim&
-\lambda'\int d\tau dy\ \hat{\sigma}^3(y,\tau)
\langle\partial_x\hat{\Phi}(x,0)\sin\big(\sqrt{\pi}\hat{\Phi}(y,\tau)\big)\rangle_{\Phi}
+\dots\nn  
&=&A\partial_x\hat{\sigma}^3(x)+\dots\,.
\eea
The leading contribution to the low-energy projection of $\hat{n}^z_j$
occurs at order $\hat{O}(\lambda')^0$ of our procedure and gives
\bea
\hat{n}^z_j\Bigl|_{\rm
  low}&\sim&B'\langle\cos\big(\sqrt{\pi}\hat{\Phi(x)}\big)\rangle_\Phi\ \hat{\sigma}^3(x)+\dots\nn   
&=&B\hat{\sigma}^3(x)+\dots\,.
\eea
\subsection{Integrating out the fermionic degrees of freedom}
This case pertains to the transition line between the D-LD and D-H
phases. Here we have
\be
   \hat{O}\Bigl|_{\rm low}=\int {\cal D}\hat{R}_3{\cal D}\hat{L}_3
   \ e^{-S_3-S_{\rm    int}}\hat{O}=\langle\hat{O}\rangle_{3}
-\langle S_{\rm int}\hat{O}\rangle_{3}+\dots\ ,
\ee
where $\langle\rangle_{3}$ denotes the average with respect to the
Majorana action $S_3$. On the transition line we have
$m_3>0$ which implies
\be
\langle\hat{\mu}^3(x)\rangle_3\neq 0.
\label{mu3}
\ee
An immediate consequence of \fr{mu3} is that
\be
\hat{n}^x_j\Bigl|_{\rm low}\sim \cos\big(\sqrt{\pi}\hat{\Theta}(x)\big)\langle \hat{\mu}^3(x)\rangle_3+\dots\,.
\ee
The low-energy projections of other operators can be worked out as
before
\bea
\hat{n}^z_j\Bigl|_{\rm low}&\sim&-\lambda'B'\int dy
d\tau\ \langle\hat{\sigma}^3(x,0)\hat{\sigma}^3(y,\tau)\rangle_3
\sin\big(\sqrt{\pi}\hat{\Phi}(x,0)\big)
\cos\big(\sqrt{\pi}\hat{\Phi}(y,\tau)\big)\nn
&=& A_z\sin\big(\sqrt{4\pi}\hat{\Phi}(x)\big)+\dots
\eea
Here we have used that
\be
\langle\hat{\sigma}^3(x,0)\hat{\sigma}^3(y,\tau)\rangle_3\propto e^{-r/\zeta}\ ,
\ee
which permits us to employ operator product expansions in the bosonic
sector. The projection of the dimerization operator is
\bea
\hat{D}_j\Bigl|_{\rm low}&\sim&
-\lambda'\int d\tau dy\ \langle\hat{\sigma}^3(x)\hat{\sigma}^3(y,\tau)\rangle_3
\sin\big(\sqrt{\pi}\hat{\Phi}(x,0)\big)
\sin\big(\sqrt{\pi}\hat{\Phi}(y,\tau)\big)+\dots\nn
&=& \langle \hat{d}\rangle+ D \cos\big(\sqrt{4\pi}\hat{\Phi}\big)+\ldots
\eea

\section{Determination of phase boundaries}
\label{qpt-xi-cstar}
\begin{figure}[tb]
 \centering
 \includegraphics[width=0.95\columnwidth,clip]{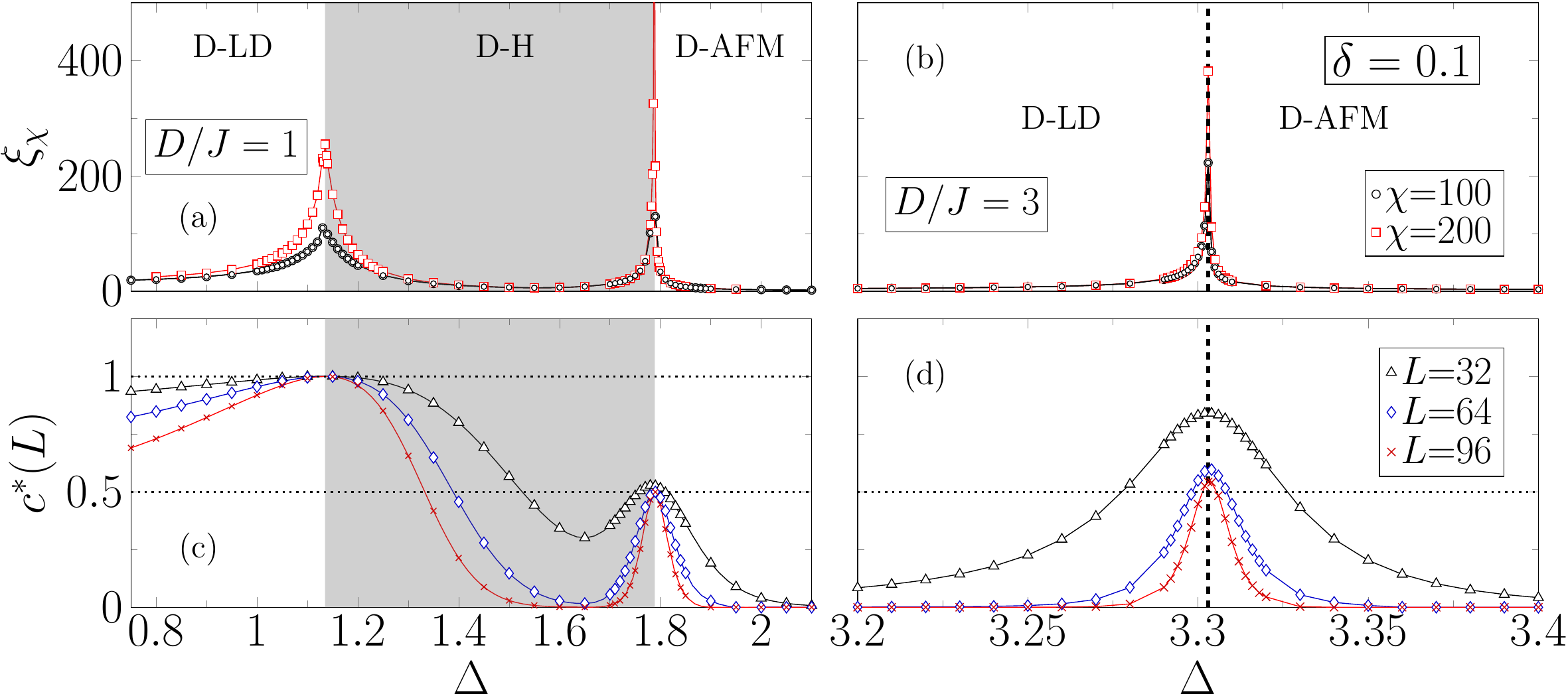}
 \caption{Correlation length $\xi_\chi$ (upper panels)
 and  central charge $c^\ast(L)$ 
 (lower panels) for fixed $D/J=1$ (left panels) and 
 $3$ (right panels) with $\delta=0.1$. 
 }
 \label{xi-cstar}
\end{figure}

In this section, we explain how  the QPT points and their universality classes 
are determined within the (i)DMRG method. 
Since the QPTs are the only points in the considered parameter region  
where the system becomes critical, they are easily obtained
by simulating the correlation length $\xi_\chi$,  as demonstrated 
in Figs.~\ref{xi-cstar}(a) and (b) for $\delta=0.1$ with fixed $D/J=1$ and $3$, 
respectively. 
The divergence of the physical correlation length at a QPT is reflected 
by a pronounced peak of $\xi_\chi$ whose height increases with the bond dimension $\chi$. 
From the peak positions for large enough $\chi$, we pinpoint 
the phase transition with an accuracy of at least three digits. 
For $D/J=1$ the transitions occur at $\Delta_{\rm c1}\simeq1.135$ and 
$\Delta_{\rm c2}\simeq1.789$ [see Fig.~\ref{xi-cstar}(a)], 
while there is only one Ising transition at $\Delta_{\rm c}\simeq 3.303$ 
[see Fig.~\ref{xi-cstar}(b)].

The central charge $c^\ast(L)$ calculated by DMRG also exhibits a peak structure
around the critical points [see Figs.~\ref{xi-cstar}(c) and (d)]. 
These peaks become more distinct with increasing system size $L$. 
From the heights of the peaks at large $L$, we obtain the central charges $c=1$ 
and $c=1/2$, which are consistent with Gaussian- and Ising-type transitions, 
respectively. Moreover, the positions of the peaks agree with the QPT points 
estimated from the correlation length.

\section{Ground-state phase diagram for strong dimerization}
\begin{figure}[tb]
 \centering
 \includegraphics[width=0.75\columnwidth,clip]{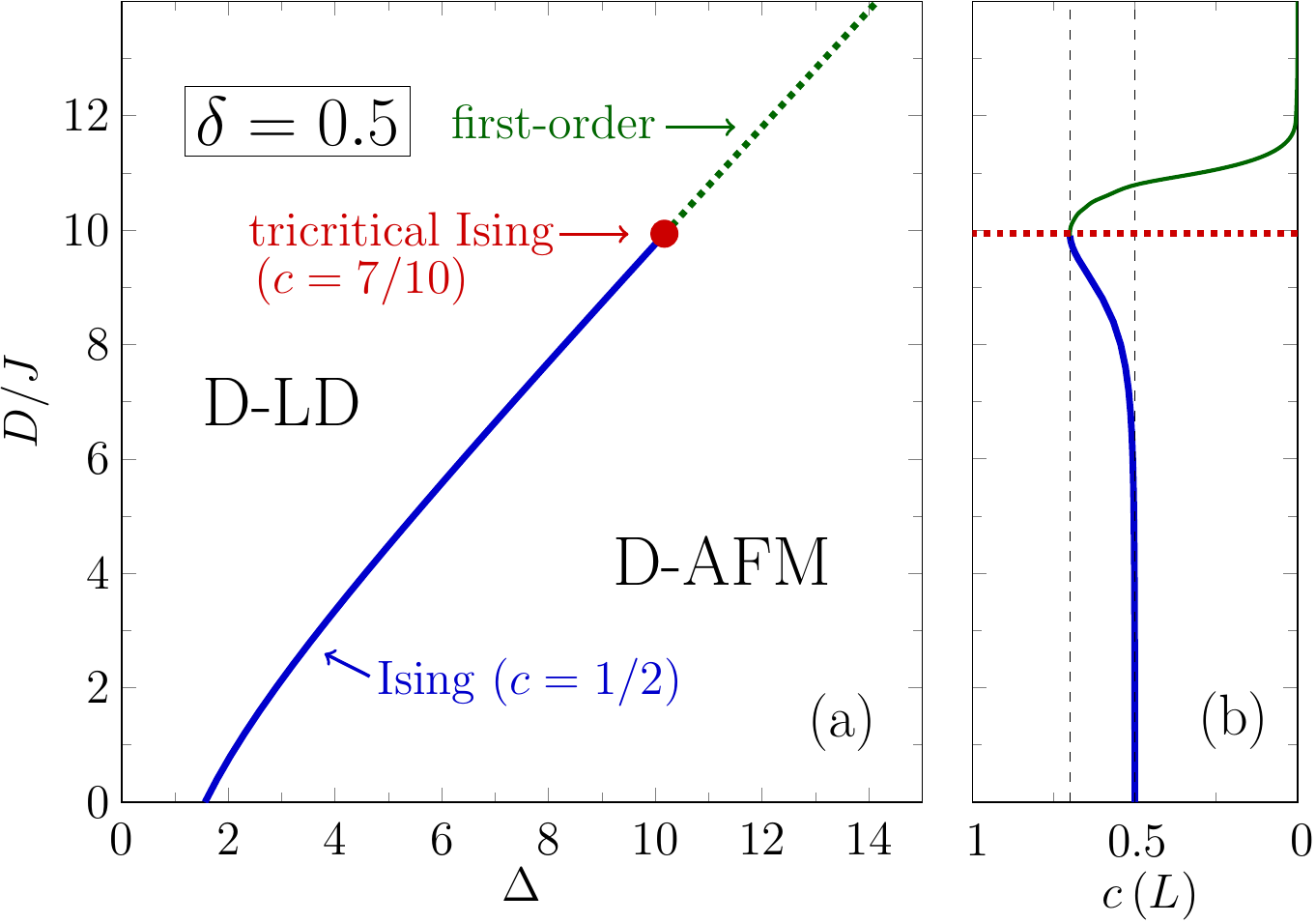}
 \caption{(a) Phase diagram of the model~\eqref{model} for $\delta=0.5$.
 D-LD{\trans}D-AFM phase boundary of the continuous Ising transition 
 terminates at a tricritical Ising point. 
 Beyond this point, the QPT becomes first order.
 (b) Central charge $c(L)$ 
 on the D-LD{\trans}D-AFM phase boundaries obtained numerically 
 for $L=128$ and periodic boundary conditions.}
 \label{pd-delta0c5}
\end{figure}
With increasing dimerization the D-H phase is reduced, and 
it disappears for $\delta\gtrsim0.26$~\cite{KN97} if we limit
ourselves to the parameter region $J>0$ and $\delta>0$. 
Figure~\ref{pd-delta0c5}(a) for $\delta=0.5$ demonstrates 
such a situation consisting of only D-LD and D-AFM phases, 
separated by continuous and first-order transition lines. 
At the meeting of these transition lines the numerically 
obtained central charge indicates $c=7/10$ [Fig.~\ref{pd-delta0c5}(b)], 
suggesting that this point belongs to the tricritical Ising universality class.


\end{document}